# ISOSPIN DEPENDENCE OF INCOMPLETE FUSION REACTIONS AT 25 MEV/A

F.Amorini[a,c], G.Cardella[b]*, G.Giuliani[c], M.Papa[b], C.Agodi[a], R.Alba[a], A.Anzalone[a], I.Berceanu[g], S.Cavallaro[a,c], M.B.Chatterjee[d], R.Coniglione[a], E.De Filippo[b], A.Di Pietro[a], E.Geraci[a,c], L.Grassi[b,c], A.Grzeszczuk[f], P.Figuera[a], E.La Guidara[b,c], G.Lanzalone[b,c], N.Le Neindre[e], I.Lombardo[a,c], C.Maiolino[a], A.Pagano[b], S.Pirrone[b], G.Politi[b,c], A.Pop[g], F.Porto[a,c], F.Rizzo[a,c], P.Russotto[a,c], D.Santonocito[a], P.Sapienza[a], G.Verde[b]

    a) INFN Laboratori Nazionali del Sud, Catania, Italy
    b) INFN, Sezione di Catania
    c) Dipartimento di Fisica e Astronomia Università di Catania
    d) Saha Institute of Nuclear Physics, Kolkata, India
    e) Institut de Physique Nucleaire d'Orsay, CNRS-IN2P3, Orsay France
    f) Institute of Physics, University of Silesia, Katowice, Poland
    g) National Institute for Physics and Nuclear Engineering "Horia Hulubei", Bucharest, Romania
    * corresponding author

**ABSTRACT**

$^{40}$Ca+$^{40,48}$Ca,$^{46}$Ti reactions at 25 MeV/A have been studied using the 4π CHIMERA detector. An isospin effect on the competition between incomplete fusion and dissipative binary reaction mechanisms has been observed. The probability of producing a compound system is observed to be lower in the case of N≈Z colliding systems as compared to the case of reactions induced on the more neutron rich $^{48}$Ca target. Predictions based on CoMD-II calculations show that the competition between fusion-like and dissipative reactions, for the selected centrality, can strongly constraint the parameterization of symmetry energy and its density dependence in the nuclear equation of state.

**Pacs:** 21.65.Ef, 21.65.Mn, 25.70.Jj, 25.70.Lm

Collisions between heavy ions with different neutron-proton asymmetries offer a unique opportunity to study the equation of state (EOS) of asymmetric nuclear matter [1-4]. Accessing the density dependence of the symmetry energy has recently attracted the interest of the community due to its implications in both nuclear physics and astrophysics of neutron stars [2,5,6]. The isotopic composition of fragments produced in multifragmentation phenomena is being extensively studied at intermediate beam energies (E/A=20-100 MeV) bearing important information on the symmetry energy [7]. One aspect not yet fully investigated is represented by the effect of the isospin asymmetry on the fate of hot nuclear systems. The combined effects of the symmetry energy and of the repulsive Coulomb interaction can significantly affect the reaction mechanism and the rate of production of hot compound nuclei in heavy-ion reactions at low and intermediate energies [8-9]. The isospin N/Z-asymmetry can also play an important role in opening different decay channels for a hot nuclear system once this has been produced. In this respect, the limiting temperature of a nucleus is expected to depend on both its mass and its isotopic composition [10-12]. Experimentally, the observation of small isotopic effects on the temperatures of projectile spectators in relativistic heavy-ion collisions have been interpreted as a signal of no isospin dependence of nuclear limiting temperatures [13]. At lower beam energies, a mass and N/Z-asymmetry dependence in limiting temperatures have been explored by studying the population of the Giant Dipole Resonance (GDR) at high excitation energies [14]. A small difference in the limiting GDR excitation energy has been observed when comparing a symmetric N~Z system to a neutron rich system [15]. All these findings stimulate attempts to link N/Z effects on measured observables to the nuclear symmetry energy and its density dependence in the equation of state.

In this work we explore isospin effects in heavy residue (HR) remnants produced in incomplete fusion reactions between projectile and targets with different N/Z asymmetries. The results on the competition between incomplete fusion and binary dissipative mechanisms are compared to simulations performed with a microscopic model providing constraints to the density dependence of the symmetry energy.

The experiment was performed at the INFN Laboratori Nazionali del Sud (LNS) with $^{40}$Ca beams at 25 MeV/A delivered by the LNS Superconducting Cyclotron. Reactions on self-supporting, metallic form, $^{48,40}$Ca and $^{46}$Ti isotopically enriched targets have been studied. The thicknesses of the targets were, respectively, 2.87, 1.24 and 1.06 mg/cm$^2$. Produced fragments were detected by the CHIMERA [16] 4π multi-detector, consisting of 1192 silicon-CsI(Tl) telescopes, arranged in 26 rings covering 95% of 4π from 1° to 176°. During the measurements, the most forward rings, up to 4.5°, were covered. The event trigger conditions of the experiment required at least 3 silicon detectors fired by fragments with charge Z>1 in order to remove the most peripheral collisions from the analysis. Detected particles were identified in charge and mass with standard ΔE-E and Fast-Slow identification techniques [17]. Furthermore, Time of Flight (TOF) measurements provided the particle velocity [18]. Such TOF measurements were performed using the Cyclotron radiofrequency as reference time and the silicon detectors as stop signals. The standard overall time resolution obtained with this method is of the order of 1 ns, mostly due to beam time characteristics. Combining the TOF and energy measurement, the mass of particles stopped in silicon detectors was evaluated. In the most forward rings the flight distance of 2-3m gives a mass resolution δM/M~1/25-1/30. Due also to systematic errors caused by the different rise-time of signals in silicon detectors [19], a total error of ±5% has been estimated for masses A~50, i.e. the typical mass of HR fragments relevant to the analysis presented in this work.

Due to handling difficulties, $^{40}$Ca and $^{48}$Ca targets underwent oxidation. However, uninteresting reactions induced by the beam on oxygen contaminants in the targets have been isolated and removed from the data analysis by studying only complete events. These events were defined by requiring the sum of the charges of all detected fragments, $\Sigma Z_i$, to be larger than 80% of the total charge of the colliding system (projectile and target), i.e. $\Sigma Z_i$>32. In order to better select complete events, we required the sum of all fragment momenta, $\Sigma P_i$, to be larger than 70% of the total momentum in the entrance channel. Furthermore, events produced by a possible pile-up of two or more reactions were excluded by requiring $\Sigma Z_i$ to be smaller than 40, for Ca targets, and smaller than 42, for Ti targets.

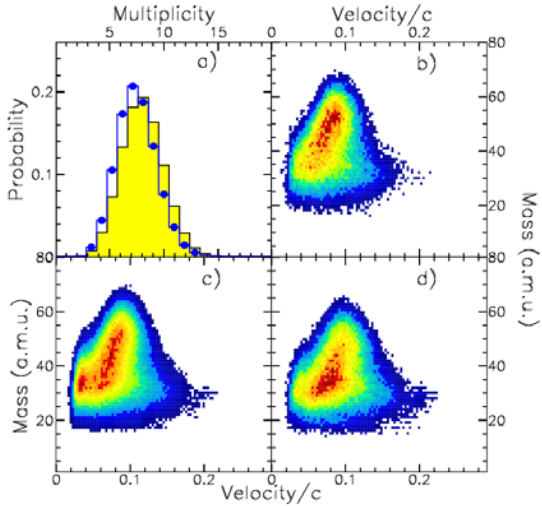

Fig.1 a) Charged particle multiplicity in the reactions $^{40}Ca+^{48}Ca$ (dot histogram) and $^{40}Ca+^{46}Ti$ (shaded area histogram) for complete events (see text). b) Mass $m_1$ versus velocity $v_1$ of the largest fragment detected in the $^{40}Ca+^{48}Ca$ reaction (see text for the selection used). c) same as b) but for $^{40}Ca+^{46}Ti$ reaction. d) same as b) but for $^{40}Ca+^{40}Ca$ reactions.

Charged particle multiplicity distributions are shown in fig.1a for the $^{40}Ca+^{48}Ca$ reactions (full dots) and for $^{40}Ca+^{46}Ti$ (shaded area histogram). The distribution obtained with N/Z symmetric $^{40}Ca$ targets, being very similar to that observed with $^{46}Ti$ targets, is not shown in fig.1a. The maximum in the multiplicity distributions is shifted back by one unit in the case of $^{48}Ca$ targets, probably due to the larger neutron emission probability. Taking into account this observation, in order to minimize contributions from very peripheral reactions, only events with a multiplicity of at least 5 charged particles, in the case of $^{48}Ca$ target, and 6 charged particles, in the case of $^{46}Ti$ and $^{40}Ca$ targets, have been selected. In order to better isolate events where incomplete fusion occurs, we required the velocity of the second or third heaviest detected fragment to be larger than 0.13c. In fact, in such incomplete fusion reactions, the portion of the projectile that does not fuse with the target is expected to move with a velocity close to that of the projectile. We note that our selections strongly suppress events in which only a part of the target fuses and the remaining part behaves like a spectator [20]; target spectators are in fact not detected and do not contribute to the $\Sigma Z_i$. Fig.1 b, c, d show the correlations between the mass, $m_1$, and the velocity, $v_1$, of the largest detected fragment in the reactions $^{40}Ca + ^{48}Ca$, $^{40}Ca + ^{46}Ti$ and $^{40}Ca+^{40}Ca$ respectively. The distribution on Fig.1b is peaked at velocities $v_1$~0.09c, close to the centre of mass velocity of the $^{40}Ca+^{48}Ca$ reaction ($V_{cm}$=0.105c), and at masses $m_1$~50 a.m.u.. These large fragments are the HR remnants of incomplete fusion reactions. The distributions obtained for the other two reactions (Fig.1c, 1d) regardless their similarities, display some marked differences. Target-like nuclei, with a velocity ≈0.04c, are produced more copiously in the case of a $^{46}Ti$ target (fig.1c) than in the case of a $^{40}Ca$ target (fig.1d). These heavy and slow target-like fragments are mostly stopped in thicker $^{48}Ca$ targets, therefore they are not observed when studying $^{40}Ca+^{48}Ca$. The fragment mass distributions are clearly different for the three studied reactions. Indeed, these distributions are peaked to much lower masses in the case of $^{40}Ca$ targets (Fig.1d).

In order to further explore mass correlations, we plot on Fig. 2a the difference between the masses of the two largest fragments normalized to the total initial mass of the system, $\Delta M_{nor} = (m_1-m_2) / m_{tot}$ (with $m_{tot} = m_{target} + m_{projectile}$), for $^{48}Ca$ (dots), $^{46}Ti$ (shaded-area histogram) and $^{40}Ca$ (empty histogram) targets. In order to exclude target-like contributions only velocities greater than 0.04c have been taken into account. The $\Delta M_{nor}$ observable allows one to disentangle dissipative binary collisions and/or fusion-fission events, characterized by two large fragments with similar masses (small $\Delta M_{nor}$ values), from fusion-like events producing heavy residues (higher $\Delta M_{nor}$ values, close to unity). An enhancement at $\Delta M_{nor}$~0.5 is observed in the case of reactions induced on the $^{48}Ca$ target. The presence of the enhancement is confirmed in Fig.2b where we plot the $m_1/m_{tot}$ spectrum for events, mainly produced in fusion reactions, selected with the condition $m_2$<10. On the other hand, in the N≈Z symmetric reaction $^{40}Ca+^{46}Ti$, we observe an enhancement of the $\Delta M_{nor}$ distribution around $\Delta M_{nor}$~0.2.

The larger probability of producing heavy-residues observed in $^{40}Ca+^{48}Ca$ reactions may be connected to the larger neutron content of $^{48}Ca$.

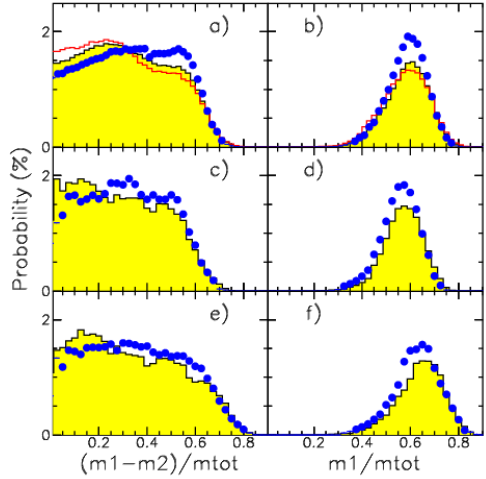

*Fig.2 a) Probability plots of the difference between the mass of the two larger fragments emitted in the investigated reactions, normalized to the total initial mass of the system. Points, shaded area and empty histograms are respectively for reactions on $^{48}Ca$, $^{46}Ti$ and $^{40}Ca$ targets. b) Experimental $m_1/m_{tot}$ spectra for incomplete fusion events. c,d) CoMDII+ GEMINI calculations for $^{48}Ca$ and $^{46}Ti$ targets. e),f) as c,d but without the last GEMINI stage.*

In fact the neutron excess of the $^{40}Ca+^{48}Ca$ system pushes the formed hot compound nucleus closer to the stability valley, on the contrary the intermediate compound systems formed with the other two proton-rich targets are much closer to the proton drip line. This simple consideration must be obviously corrected by considering that pre-equilibrium emission of light particles can modify the isotopic content of the compound system. We will better see this point later in the text.

In order to investigate the mechanisms responsible of the observed effects, we compared our experimental results to calculations performed with the CoMD-II (Constrained Molecular Dynamics II) model [21]. In this model, the dependence of the symmetry interaction per nucleon on the total overlap integral, $s$, between the wave-packets is characterized by a form factor $F(s/s_{g.s})$ (the label g.s corresponds to the ground state configuration). This form factor can be expressed, for compact configurations, as: $F(s/s_{g.s})= s/s_{g.s} F'(s/s_{g.s})$ [22]. Depending on the specific choice of $F'(s/s_{g.s})$, one can select a different stiffness of the density dependence of the symmetry energy. Specifically, we have used: $F'(s/s_{g.s})=2(s/s_{g.s}) / (1+s/s_{g.s})$ (Stiff1); $F'(s/s_{g.s})=1$ (Stiff2); $F'(s/s_{g.s})=(s/s_{g.s})^{-1/2}$ (Soft). The $\Delta M$ and $m_1/m_{tot}$ distributions predicted by CoMD-II are represented in Fig. 2e and 2f, respectively for $^{46}Ti$ and $^{48}Ca$ targets (same symbols of Fig.2a,b; being experimental results on $^{40}Ca$ targets very similar to the $^{46}Ti$ case, we did not perform calculations on such target). These calculations are performed using the above reported Stiff2 parameterization for the symmetry term. The dynamical evolution of the system has been determined up to 600 fm/c. In order to take the effects induced by secondary decays into account, the statistical decay of the excited primary fragments produced at this stage of the CoMD-II calculations has been computed with the GEMINI code [23]. The ensemble of the simulated events have been finally filtered through the angular coverage and detector efficiency of CHIMERA. Moreover, a selection on the events reflecting the main criteria of the data analysis have been also included. Fig. 2c-d show that the trend of the data is satisfactory reproduced by CoMD-II calculations.

The effect of preequilibrium emission on the N/Z ratio of the formed compound system can be simply seen in the CoMD-II by calculating the charge/mass asymmetry ($\beta=(N-Z)/A$) of the produced HR. After a time of 600 fm/c, due to the interplay between Coulomb and symmetry interactions, this asymmetry resulted to be, on average, $\beta\sim0.05$ and 0.1 for the $^{40}Ca+^{46}Ti$ and $^{40}Ca+^{48}Ca$ reactions, respectively, to be compared to the initial compound system asymmetry values of 0.02 and 0.09. Therefore, even if preequilibrium emission, during the dynamical stage, pushes the system toward the stability valley, the two colliding systems still maintain a quite different asymmetry between them.

The comparison between the results obtained with CoMD-II only (Fig. 2e,f) and CoMD-II+GEMINI (Fig. 2c,d) simulations clearly suggests that the suppression of incomplete fusion events in N/Z symmetric systems is mostly determined by the dynamical stage of the reaction and not affected by statistical secondary decays.

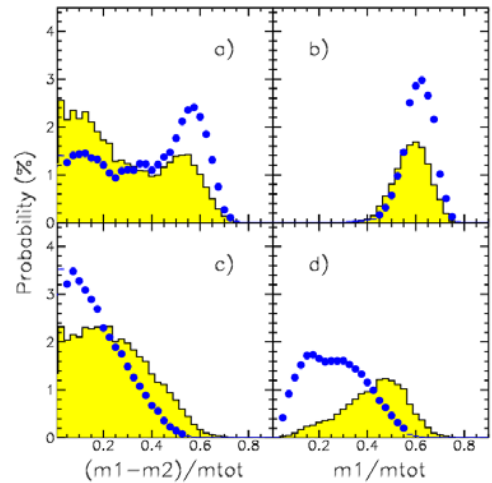

*Fig.3 CoMDII+GEMINI calculations as for fig.3c,d. a,b) stiff1 parameterization; c,d) soft parameterization.*

On Fig. 3 we represent the same CoMD-II distributions plotted in Fig. 2c,d but using different parameterizations of the symmetry potential (panels a and b: Stiff1 parameterization; panels c and d: Soft parameterization). In the *Stiff1* and *Stiff2* cases, the many-body correlations of CoMD-II generate an attractive interaction as the nuclear density grows, inducing a higher yield of HR. The Soft symmetry potential maintains instead a repulsive contribution producing a fragmentation of the source. The mentioned many-body correlations [22] affect significantly the strength of the symmetry interaction. This explains the large changes observed in the mass distributions for the three different parameterizations (see Figs. 2,3). CoMD-II calculations show that the observables experimentally explored on Fig. 2 can be used to probe the symmetry potential in nuclear matter. From the comparison to experimental data, a quantitative determination of the degree of stiffness of the equation of state, describing the dependence of the symmetry interaction on the total overlap integral, $s$, can be obtained by using a simple interpolation procedure. As it was mentioned above, in the CoMD-II model the symmetry potential is characterized by a form factor, $F(s/s_{g.s}) = s/s_{g.s}F'(s/s_{g.s})$, depending on the total overlap integral s from compact configurations. We reasonably assumed that the dynamical evolution of the system strongly depends on the value of the maximum density overlap, $s_{max}$, achieved during the first 100 fm/c. At this stage of the dynamical evolution of the system, we can express the $F'(s/s_{g.s})$ as linearly depending on the quantity $(s-s_{g.s})/s_{g.s}$. This quantity is relatively small at our energies ( ≈10-15%), so that one can approximate $F'(s/s_{g.s})≈1+\delta\times(s-s_{g.s})/s_{g.s}$. The form factors used in the case of Stiff1, Stiff2 and Soft symmetry potentials correspond, respectively, to $\delta=0.5$, $\delta=0$ and $\delta=-0.5$. Determining the value of $\delta$ that produces the best description of the data would allow us to quantitatively determine the stiffness, $\gamma=1+\delta$, of the symmetry potential $F(s/s_{gs})$. In order to accomplish this, we choose three characteristic points on the probability distributions on Figs. 2 and 3, i.e. $P_1=P(\Delta M/m_{tot}=0.5)$, $P_2=P(\Delta M/m_{tot}=0.2)$ and $P_3=P(M_1/m_{tot}=0.5)$. We calculate the values of $P_1$, $P_2$ and $P_3$ obtained from CoMD-II calculations performed for all reaction systems and with the Stiff1, Stiff2 and Soft symmetry potential parameterizations. As a first approximation we can assume a quadratic dependence of $P_j$ (i=1,2,3) values on $\delta$, i.e. $P_j=A_j+B_j*\delta+C_j*\delta^2$. Once the $A_j$, $B_j$, $C_j$ coefficients are determined, one can obtain what should be the $\delta=\delta^{exp}$ value that best reproduces the experimental probabilities $P_i^{exp}$ (i=1,2,3) in each projectile-target combination. We deduce an average $\delta$ value, $<\delta^{exp}>= 0.1\pm0.2$ and therefore an average value of $\gamma$ $<\gamma>=1.1\pm0.2$. This value characterizes quantitatively the symmetry interaction parameterization that best reproduces the experimentally observed isospin effect on the competition between binary dissipative an incomplete fusion events.

In summary we have observed, for the first time, an isospin effect on the competition between incomplete fusion and dissipative binary reaction mechanisms at an incident energy of 25 MeV/A. This energy regime is close enough to the opening of the multi-fragmentation decay channel, where the isospin asymmetry is expected to play a relevant role in determining the limiting temperature of a nuclear system. We observe that the production of heavy remnants by an incomplete fusion mechanism is enhanced in the case of neutron rich reaction systems (with mass of the reaction partners around 40-50 amu), while binary deeply inelastic processes dominate the mechanisms for isospin symmetric N~Z systems. According to CoMD-II model calculations, the observed isotopic effect is attributed to the interplay between Coulomb and symmetry interactions. A comparison of the measured mass correlations to CoMD-II model predictions provides constraints on the stiffness of the nuclear symmetry potential that is found to be characterized by a form factor $F≈(s/s_{gs})^\gamma$ with $\gamma≈1.1\pm0.2$.

Thanks are due to the INFN Sezione di Catania and INFN LNS staff respectively for the invaluable contribution during the setting up of the detector and for the high quality of the delivered beams. Thanks are also due to M. D'Agostino for useful discussions.